\documentclass{aa}
\usepackage{graphics,latexsym,amssymb,times,psfig,amsmath}
\def\gsim{ \lower .75ex \hbox{$\sim$} \llap{\raise .27ex \hbox{$>$}} }
\def\lsim{ \lower .75ex\hbox{$\sim$} \llap{\raise .27ex \hbox{$<$}} }

\begin{document}
   \title{Cosmological constraints with GRBs: homogeneous medium
   vs
 wind density profile} 
   \author{Ghirlanda G.  \inst{1}, Ghisellini G. \inst{1}, Firmani C.  
   \inst{1,3}, Nava L. \inst{1,2}, Tavecchio F.  \inst{1}, Lazzati D. \inst{4}}
   \offprints{Giancarlo Ghirlanda \\ghirlanda@merate.mi.astro.it}
   \institute{Osservatorio Astronomico di Brera, via Bianchi 46,
   Merate Italy \and Univ. di Milano--Bicocca, P.za della Scienza
   3, I--20126, Milano, Italy. \and Instituto de Astronom\'{\i}a,
   U.N.A.M., A.P. 70-264, 04510, M\'exico, D.F., M\'exico \and
   JILA, University of Colorado, Boulder, CO 80309-0440, USA}
   \date{Received ... / Accepted ...} \titlerunning{GRB jet
   models and cosmology}
 \authorrunning{G. Ghirlanda et al.}

\abstract{ We present the constraints on the cosmological parameters
obtained with the $E_{\rm peak}$--$E_{\gamma}$ correlation found with
the most recent sample of 19 GRBs with spectroscopically measured
redshift and well determined prompt emission spectral and afterglow
parameters. We compare our results obtained in the two possible
uniform jet scenarios, i.e. assuming a homogeneous density profile
(HM) or a wind density profile (WM) for the circumburst medium. Better
constraints on $\Omega_{M}$ and $\Omega_{\Lambda}$ are obtained with
the (tighter) $E_{\rm peak}$--$E_{\gamma}$ correlation derived in the
wind density scenario. We explore the improvements to the constraints
of the cosmological parameters that could be reached with a large
sample, $\sim$ 150 GRBs, in the future. We study the possibility to
calibrate the slope of these correlations. Our optimization analysis
suggests that $\sim 12$ GRBs with redshift $z\in(0.9,1.1)$ can be used
to calibrate the $E_{\rm peak}$--$E_{\gamma}$ with a precision better
than 1\%.  The same precision is expected for the same number of
bursts with $z\in(0.45,0.75)$. This result suggests that we do not
necessarily need a large sample of low $z$ GRBs for calibrating the
slope of these correlations.
 
  }
\maketitle

\section{Introduction}
Gamma Ray Bursts (GRBs) are presently detected out to very high
redshifts (the new limit being GRB~050904 at $z=6.29$, Kawai et
al. 2005) and this makes them extremely attractive for observational
cosmology. They might have profound impact on (i) the study of the
epoch of reionization, (ii) the characterization of the properties of
the cosmic intergalactic medium, (iii) the description of the geometry
of the Universe and (vi) the investigation of the nature of Dark
Energy.
 
GRBs have recently been discussed (Ghirlanda et al. 2004b,
Firmani et al.  2005, Lamb et al. 2005) as a possible new class of
``standard candles'' to be used to constrain the Universe models
(Ghirlanda et al. 2004b, Firmani et al. 2005, Liang \& Zhang
2005). What ``standardize'' the GRB energetics is a tight relation
between their rest frame collimation corrected energy $E_{\gamma}$ and
the peak energy $E_{\rm peak}$ of their $\nu F_{\nu}$ prompt emission
spectra (Ghirlanda et al. 2004).
 
In the ``standard'' GRB scenario
the collimation corrected energy is
 defined as $E_{\gamma}=E_{\rm
\gamma, iso}(1-\cos\theta_{\rm j})$, where
 $E_{\rm \gamma, iso}$ is
the isotropic equivalent energy and $\theta_{\rm j}$ is the jet
opening angle. This parameter can be derived from $t_{\rm j}$,
i.e. the time when the afterglow light curve steepens, in two
different scenarios: (a) assuming that the circumburst medium is
homogeneous (HM) or (b) assuming a stratified density profile
produced, for instance, by the wind of the GRB progenitor (WM). In
both cases the jet is assumed to be uniform.
 
Originally the
$E_{\rm peak}$--$E_{\gamma}$ correlation was derived in the HM case
with 15 GRBs (Ghirlanda et al. 2004) and resulted in a small scattered
correlation ($E_{\rm peak}\propto E_{\gamma}^{0.7}$ with $\sigma\sim
0.15$) which was used to derive interesting (though shallow)
constraints on the cosmological parameters $(\Omega_{\rm
M},\Omega_{\Lambda})$ and on the equation of state parameters of the
Dark Energy (Firmani et al. 2005). 
 
Recently Nava et al. 2005 (N05) derived the $E_{\rm peak}$--$E_{\gamma}$ 
correlation in the WM case.
This correlation is less scattered, i.e.  $\sigma=0.08$, than in the HM 
case and it is linear, i.e. $E_{\rm peak}\propto E_{\gamma}$.  N05
discussed its implications for the understanding of the dynamics and
radiative processes of GRBs. Although the $E_{\rm peak}$--$E_{\gamma}$
correlations (both in the HM and WM case) are model dependent, their
consistency with the completely empirical correlation between 
$E_{\rm peak}$, $E_{\gamma}$ and $t_{\rm j}$ found by Liang \& Zhang 
(2005 - hereafter LZ05), suggests that the model parameters are not much 
dispersed.  It is therefore worth to derive the cosmological constraints 
with the Ghirlanda correlation in both the HM and WM scenario and compare 
the results.
 
In order to use GRBs as cosmological tools, through the
above correlations, three foundamental parameters, $E_{\rm peak}$,
$E_{\gamma}$ and $t_{\rm jet}$, should be accurately measured or
inferred.  This requirement also applies to the empirical correlation
of LZ05. Therefore only a limited number of GRBs, i.e. 19 out of
$\sim$ 70 (up to Nov. 2005) with measured $z$, can be used as
standard candles.  It also appears that the limited energy range of
BAT onboard Swift (15--150 keV) allows to constrain only with moderate
accuracy the $E_{\rm peak}$ of particularly bright--soft
bursts. However, given the perspective of the cosmological
investigation through GRBs, it is worth exploring the power of using
GRBs as cosmological probes.
 
Another still open issue related to the use of GRBs as standard candles 
is the so called ``circularity problem'' (see Ghisellini et al. 2005).  
This is due to the fact that
the small number of GRBs with spectroscopic measured redshifts and
their wide dispersion in $z$ does not allow to calibrate the Ghirlanda
correlation, which, in turn, depends on the cosmological parameters
that we want to constrain.  To the aim of calibrating this
spectral--energy correlation, GRBs at low redshift are required.  In
fact for $z<0.1$ the difference in the luminosity distance computed
for different choices of the cosmological models [for $\Omega_{\rm
M},\Omega_{\Lambda}\in(0,1)$] is less than 8\%.  However, if (long)
GRBs are produced by the death of massive stars, they should roughly
follow the cosmic star formation history (SFR) and we should expect
that the rate of low redshift events is quite small at $z<0.1$.
Instead, a considerably large number of GRBs with $z>1$ should be
collected in the next years by presently flying instruments (Swift and
Hete--II). At such large redshifts, the cosmological models starts to
play an important role.  However, if it will be possible to have a
sufficient number of GRBs with a similar redshift, it might still be
possible to calibrate the slope of these correlations with high
redshift GRBs.
\begin{figure}
\centerline{\psfig{figure=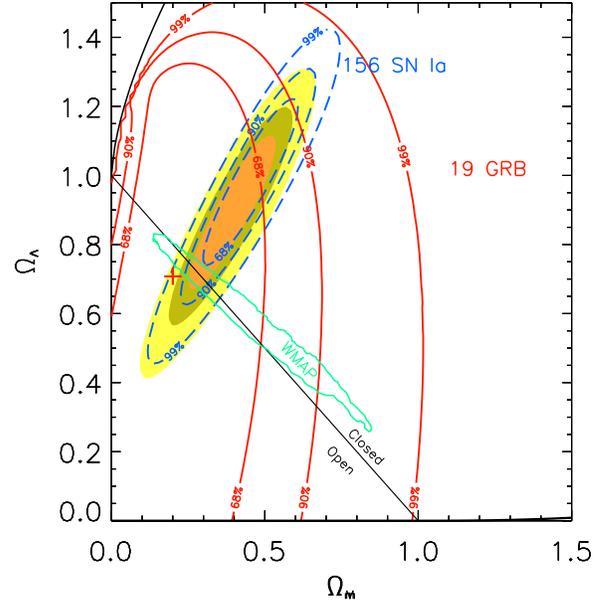,width=9cm}}
\vskip -0.2 true cm
\caption{
Constraints on the cosmological parameters
$\Omega_{\rm M}$, $\Omega_{\Lambda}$ obtained with the updated sample of 19
GRBs presented in Tab. 1 of N05 (to which GRB~051022 was added) in the homogeneous 
density case (HM). The solid (red) contours, obtained with the 19 GRBs alone,
represent the 68.3\%, 90\% and 99\% confidence regions. The center of
these contours (red cross) corresponds to a minimum
$\chi^{2}=15.25/17$ dof and has $\Omega_{M}=0.23$ and
$\Omega_{\Lambda}=0.81$. The contours obtained with the 156 SN Ia of
the ``Gold'' sample of Riess et al. 2004 are shown by the dashed
(blue) lines. The joint GRB+SN constraints are represented by the
shaded contours. We also show the 90\% confidence contours obtained
with the WMAP data (from Spergel et al. 2003).
}
\label{cosmo_new}
\end{figure}
\begin{figure}
\centerline{\psfig{figure=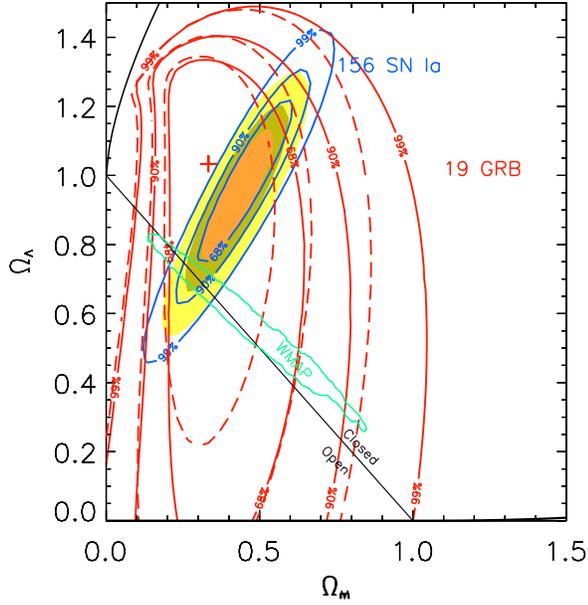,width=9cm}}
\vskip -0.2 true cm
\caption{
Constraints on the cosmological parameters
$\Omega_{\rm M}$, $\Omega_{\Lambda}$ obtained with the updated sample of 19
GRBs presented in Tab. 1 of N05 (to which GRB~051022 was added) in the wind 
density profile case (WM). The solid (red) contours, obtained with the 19 GRBs 
alone, represent the 68.3\%, 90\% and 99\% confidence regions on the pair of
cosmological parameters. The contours obtained with the 156 SN Ia of
the ``Gold'' sample of Riess et al. 2004 are shown by the long--dashed
(blue) lines. The joint GRB+SN constraints are represented by the
shaded contours. We also report (dashed--red contours) the constraints
obtained by assuming a fixed (i.e. cosmology invariant) linear
correlation $E_{\rm peak}\propto E_{\gamma}$. }
\label{cosmo_wind} 
\end{figure}

The aim of this work is to find the cosmological constraints by
using the $E_{\rm peak}-E_{\gamma}$ correlation in the HM and WM
case (Sec. 3) and to describe the level of precision that could be
achieved with a large population of bursts (Sec. 4). Finally, we
study the possibility to calibrate the slope of these relations with
GRB samples at intermediate redshifts (Sec. 5).
 
\section{The $E_{\rm peak}$-$E_{\gamma}$ correlation in the HM and WM case}

The jet opening angle $\theta_{\rm j}$ in the two HM or WM density profile scenarios is:
\begin{equation}
\theta_{\rm j}=0.161 \left({ t_{\rm j,d} \over 1+z}\right)^{3/8}
\left({n \, \eta_{\gamma}\over E_{\rm \gamma,iso,52}}\right)^{1/8}\,
\, \, \, \, {\rm HM}
\label{theta}
\end{equation}
\begin{equation}
\theta_{\rm j, w}\, = \, 0.2016 \, \left( {t_{\rm j,d} \over
1+z}\right)^{1/4} \left( { \eta_\gamma\ A_* \over E_{\rm
\gamma,iso,52}}\right)^{1/4}\, \, \, {\rm WM}
\label{thetaw}
\end{equation}
where $n$ is the constant circumburst density in the HM case
(Sari et al. 1999) and $A_*$ is the normalization of the density profile
$n(r)=5\times10^{11} A_{*}r^{-2}$ g cm$^{-3}$ in the WM case (Chevalier
\& Li 2000). 

N05 derived the $E_{\rm peak}$-$E_{\gamma}$ correlation in the HM
and
 WM case with 18 GRBs (sample updated to Sept. 2005) for which
secure
 measurements of $z$, $E_{\rm peak}$ and $t_{\rm j}$ have been
reported
 in the literature. While completing this work the Hete--II
and Konus--Wind satellites detected a new GRB (051022 - Olive et
al. 2005) for which all the observables (i.e. $z=0.8$ [Butler et
al. 2005], $E_{\rm peak}$, $t_{\rm j}$) required to compute the
$E_{\rm peak}$-$E_{\gamma}$ correlation were measured. We therefore
added this burst to the sample of 18 events of N05. The spectrum of
GRB~051022 as observed in the widest energy range by Konus--Wind
(Golenetskii et al. 2005) is fitted by a powerlaw model with an high
energy exponential cutoff: the low energy photon spectral index is
$\alpha=-1.17\pm0.04$ and the ($E F_{E}$) peak energy is $E_{\rm
peak}$=510$\pm$35 keV. The 2keV--20MeV energy fluence is
2.61$\pm$0.08$\times 10^{-4}$ erg/cm$^2$. The analysis of the X--ray
light curve has shown the presence of a possible jet break at $t_{\rm
b}=2.9\pm0.2$ days (Racusin et al. 2005). For a standard
 cosmology
with $\Omega_{M}=0.3$ and $\Omega_{\Lambda}=h=0.7$ (and assuming
typical values for the density and the wind profile normalization -
see N05), we derive $\theta_{\rm j}$(HM)=$6.37\pm0.75^\circ$ and
$\theta_{\rm j}$(WM)=$3.3\pm0.1^\circ$ for the HM and WM case.  With
the addition of this new bursts to the sample of 18 events of N05, we
find $E_{\rm peak}\propto E_{\gamma}^{0.67\pm0.04}$ (with a reduced
$\chi^2_{\rm red}=1.4$ and a gaussian scatter of the 19 GRBs around
this correlation with a $\sigma=0.1$) and $E_{\rm peak}\propto
E_{\gamma}^{1.00\pm0.06}$ (with a reduced $\chi^2_{\rm red}=1.14$ and
a gaussian scatter of the 19 GRBs around this correlation with a
$\sigma=0.08$) in the HM and WM case, respectively.

\section{Cosmological Constraints}

To the aim of constraining the cosmological parameters $\Omega_{\rm M}$
and $\Omega_{\Lambda}$ we use the $E_{\rm peak}$--$E_{\gamma}$
correlations following the Bayesian method proposed by Firmani et
al. (2005). The power of this optimized method is to circumvent the
``circularity problem'' arising from the fact that these correlations
cannot be calibrated with the present sample of GRBs (see also
Sec. 4).

\subsection{Homogeneous density}

First we assume a HM and, through Eq. 1, derive the cosmological
constraints reported in Fig. \ref{cosmo_new}.  The results, obtained
with the present sample of 19 GRBs, are fully consistent with those
obtained by Firmani et al. 2005 with the smaller (and slightly
different) sample of 15 events (Tab. 1 of GGL04). The slightly larger
contours reported in Fig. \ref{cosmo_new} are due to the changes of the
parameters (and associated uncertainties) for GRB 991216, 011211,
020124, 020405, 020813, 030226, 030328, 030329, 030429 present in both
samples (see N05 for a detailed discussion).  The joint GRB+SN fit
(shaded regions in Fig. \ref{cosmo_new}) is dominated by the small
contours determined with the 156 SN Ia of the ``Gold'' sample of Riess
et al. (2004). However, we note that the GRB+SN Ia joint fit is
consistent at the 68\% confidence level with the concordance model
$\Omega_{M},\Omega_{\Lambda}=(0.3,0.7)$.

\subsection{Wind density profile}

In the case of a wind density profile we can still derive the
constraints on the cosmological parameters adopting Eq. 2 to compute
the jet opening angle. We make the simplest assumption of
$A_{*}=1$. This choice corresponds to assuming a typical mass loss
rate $\dot M_{\rm w} =10^{-5} M_\odot$ yr$^{-1}$ and wind velocity
$v_{\rm w}=10^3$ km s$^{-1}$. We do not have any knowledge of the
uncertainty associated to the parameter $A_{*}$. However, as we want
to compare these results with those obtained in the HM case (sec. 3.1)
where a typical uncertainty was asssociated to the density parameter
$n$, we assumed a 20\% error on $A_{*}$. The slightly smaller scatter
of the Ghirlanda correlation in the case of the wind density profile
and its better reduced $\chi^2=1.14$ (for 17 dof), compared to the
homogeneous case, are responsible for the more stringent constraints 
on the 
cosmological parameters in  the $\Omega$ plane. 
Our results are reported in Fig. \ref{cosmo_wind}.  
The 68\% confindence contours obtained with GRBs are still consistent 
with the concordance model although the center of the contours corresponds 
to a quite large value of $\Omega_{\Lambda}$.

On the other hand, in the WM case the $E_{\rm peak}$-$E_{\gamma}$
correlation is linear (in the standard cosmology), and thus it has the
remarkable property to be ``Lorentz invariant¨, since both the
 $E_\gamma$ and $E_{\rm peak}$ are boosted by $\sim 2\Gamma$
when transforming from the comoving to the observer frame if we assume to view a uniform
jet within the cone defined by its aperture angle.  This property
makes this correlation easier to be interpreted theoretically (see
N05).  We have then repeated our calculation assuming that the
correlation remains linear in any cosmology.  As discussed in N05, the
linear Ghirlanda correlation obtained in the wind case is still
consistent with the empirical correlation found by LZ05 and might have
important implication for the physical interpretations of GRBs. If we
then fix the slope to 1, the only free parameter remaining is the
normalization.  In this case we can find even better constraints
(dotted red contours in Fig. \ref{cosmo_wind}).

\section{GRB sample simulation}

In order to fully appreciate the potential use of GRBs for the
cosmological investigation, we simulate a sample of bursts 
comparable in number to the ``Gold'' sample of 156 SN Ia. Similar
simulations have been presented in the literature (Xu, Dai \& Liang
2005; Liang \& Zhang 2005); however different a-priori assumptions can
be made on the properties of the simulated sample and the results are
clearly dependent on these assumptions. In particular, simulations
based on the observed parameters (Xu et al. 2005) strongly depend on
the selection effects on these quantities.

We adopt here a method which makes use of the intrinsic properties of
GRBs as described by the Ghirlanda and the Amati correlations (Amati
et al. 2002). We use the most updated version of these correlations
(N05) as found with the sample of 19 GRBs. With this sample the Amati
correlation results: 
\begin{equation}
\left({E^\prime_{\rm p} \over 100\, {\rm keV}}\right) \, =\,
(0.361\pm0.02)\,
\left({E_{\gamma,\rm iso}\over 7.6 \times 10^{52}\, 
{\rm erg}}\right)^{0.57\pm 0.02}
\label{amati}
\end{equation}
with a reduced $\chi^{2}_{\rm red}=5.22$ for 16 dof.

The assumptions of our simulated GRB sample are:
\begin{itemize}
\item we assume that GRBs have an ``isotropic energy'' function
which is described by a powerlaw $N(E_{\rm iso})\propto E_{\rm
iso}^{\delta}$ for $E_{\rm iso,min}<E_{\rm iso}< E_{\rm
iso,max}$ (e.g. Firmani et al. 2004). Further we assume that GRBs 
follow the cosmic star formation rate (SFR);  
\item we use the Amati correlation (Eq. \ref{amati}) to derive the peak
energy $E_{\rm peak}$;
\item we model the scatter of the simulated GRBs around the Amati
correlation with a gaussian distribution with $\sigma=0.3$ (which
corresponds to the present scatter of the 18 GRBs around their best
fit correlation - Eq. \ref{amati});
\item we use the Ghirlanda correlation as found with the 18 GRBs
(either Eq. 1 or Eq. 2) to calculate $E_{\gamma}$;
\item we model the scatter of the simulated GRBs around the Ghirlanda
correlation with a gaussian distribution with $\sigma=0.1$ (0.08 for
the wind case - see N05);
\item we derive the jet opening angle $\theta_{\rm j}$ and the
corresponding jet break time $t_{\rm j}$;
\item we assume that the simulated GRB spectra are described by a Band
model spectrum (Band et al. 1993) with typical low and high energy
spectral photon indices $\alpha=-1.0$ and $\beta=-2.5$ and require
that the simulated GRB fluence in the 2-400 keV energy band is above a
typical instrumental detection threshold of $\sim 10^{-7}$
erg/cm$^2$. This corresponds roughly to the present threshold of
Hete-II in the same energy band.
\end{itemize}

Following the procedure described above we built a sample of 150 GRBs
with the relevant parameters: $z$, $E_{\rm iso}$, $E_{\rm p}$,
$t_{\rm, j}$. The errors associated to these parameters are assumed
to be cosmology invariant and they are set to 10\%, 20\% and 20\% for
$E_{\rm iso}$, $E_{\rm p}$, $t_{\rm, j}$ respectively.  For our
simulation we considered the Ghirlanda correlation in the WM case and
adopted the SFR\#2 of Porciani \& Madau (2001). We modeled the GRB
intrinsic isotropic energy function with a powerlaw with $\delta=-1.3$
between two limiting energies ($10^{49}$-$10^{55}$ erg). 
This particular choice of parameters is due to the requirement that the 
distributions of the relevant quantities (shown in Fig. 3) of the simulated 
sample are consistent with the same distributions for the present sample of 19
GRBs. The sample is simulated in the standard cosmology ($\Omega_{M}=0.3$ and
$\Omega_{\Lambda}=h=0.7$). We show the distribution of $z,t_b,E_p$ and $E_{iso}$ 
for the 150 simulated  bursts in comparison with the same distributions of the 
19 GRBs in Fig. 3. We also note that by choosing a steeper energy function 
we obtain a much larger number of XRF and XRR with respect to normal GRBs.

\begin{figure}
\centerline{\psfig{figure=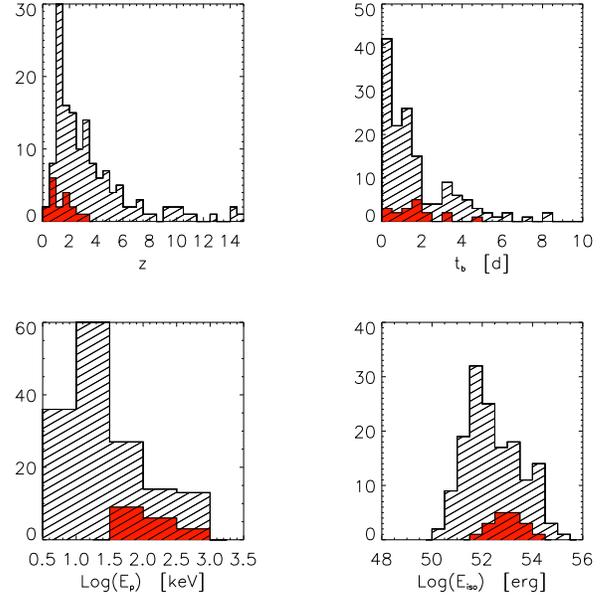,width=9cm}}
\vskip -0.2 true cm
\caption{
Distributions of redshift $z$, jet break time $t_{b}$,
observed peak energy $E_{\rm peak}^{\rm obs}$ and isotropic equivalent
energy $E_{\rm iso}$ for the 150 simulated bursts (hatched
histograms). Also show (solid filled histogram) are the distributions
of the sample of 18 GRBs of N05 used to derive the 
$E_{\rm peak}-E_{\rm iso}$ correlation. }
\label{cosmo_150} 
\end{figure}

The results obtained with the sample of 150 simulated GRBs is
presented in Fig. \ref{cosmo_150}: in this case the constraints are
comparable with those obtained with SN Ia. The minimum of the GRB
contours (cross in Fig. 4) corresponds to $\Omega_{M}=0.27$ and
$\Omega_{\Lambda}=0.72$. By comparing the 1$\sigma$ contours of GRB alone 
from Fig. 4 to the same contours (solid line) of Fig. 2 (obtained with the 19 
GRBs), we note that there is an improvement (of roughly a factor 10) with the 
sample of 150 bursts.  

It is evident the different orientation 
of the GRB contours (see Ghisellini et al. 2005) with respect to SN Ia due
to the ``topology'' of the luminosity distance as a function of the
$\Omega_{M}$-$\Omega_{\Lambda}$ parameters. Most GRBs of our simulated
sample are at $z\sim 2$, and this explains the tilt of the contours
obtained in the $\Omega$ plane (Ghisellini et al. 2005). Clearly
the simulated sample depends on the assumptions: in particular we have
no knowledge of the burst intrinsic energy function
$N(E_{\rm \gamma,iso})$. However, if we accept the hypothesis to model it
with a simple powerlaw, we can change the slope and also include the
effect of the redshift evolution. We tested the dependence of these
assumptions on the constraints reported in Fig. 4 and found that the
major effect of assuming different $\delta$ values and a $(1+z)$
evolutionary factor is to change the redshift distribution of the
simulated sample and therefore to change the orientation of the GRB
contours in Fig. 4. The same happens if we adopt, for the same choice of 
parameters reported above, a different SFR.

Further, we can also use the CMB priors. First we assume the cosmological 
constant model with the 2 CMB priors, i.e (i) $\Omega_{tot}=1$ and (ii)
$\Omega_{M}=0.14/h^{2}$. In this case the only free parameter is $h$
(or equivalently $\Omega_{M}$).We obtain the best fit values of
$\Omega_{M}=0.27\pm0.02$ and $\Omega_{\Lambda}=0.73\pm0.02$.

\begin{figure}
\centerline{\psfig{figure=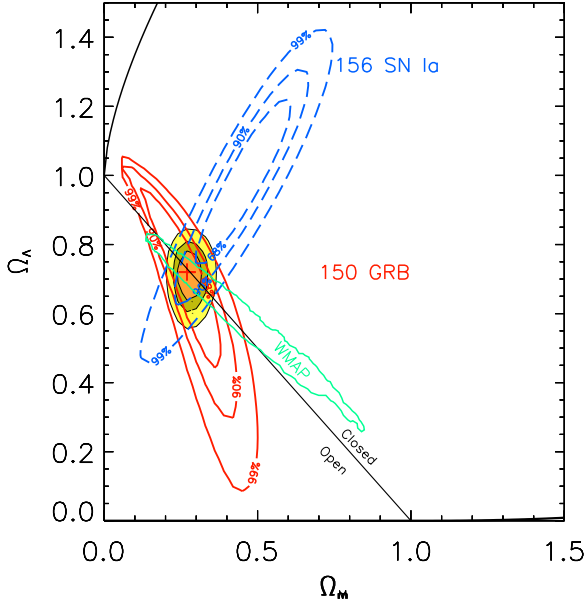,width=9cm}}
\vskip -0.2 true cm
\caption{
Wind density profile case. Constraints on the cosmological
parameters $\Omega_{\rm M}$,$\Omega_{\Lambda}$ obtained with the simulated
sample of 150 GRBs. The solid (red) contours, obtained with the 19
GRBs alone, represent the 68.3\%, 90\% and 99\% confidence regions on
the pair of cosmological parameters. The contours obtained with the
156 SN Ia of the ``Gold'' sample of Riess et al. 2004 are shown by the
dashed (blue) lines. The joint GRB+SN constraints are represented by
the shaded contours.}
\label{cosmo_150} 
\end{figure}

\subsection{Dark Energy EOS}

\begin{figure}
\centerline{\psfig{figure=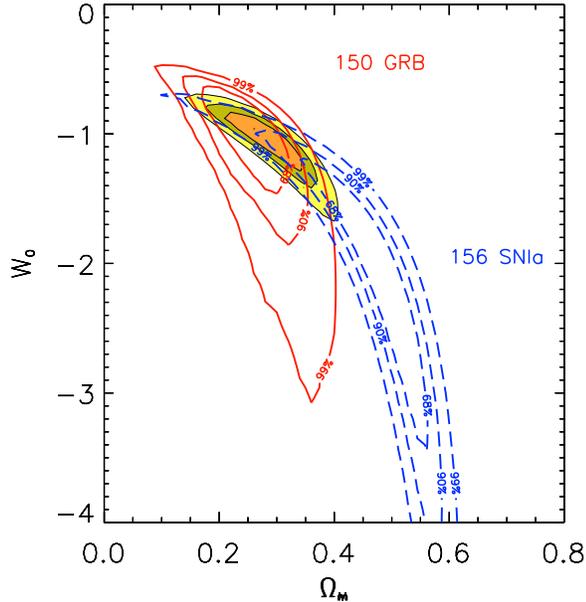,width=9cm}}
\vskip -0.2 true cm
\caption{
Constraints on the cosmological parameters
$w_{0}$, $\Omega_{\rm M}$ obtained with the 150 simulted GRBs (red
contours) compared with the same contours obtained with the 156 SNIa
of the ``Gold'' sample. A flat universe is assumed
($\Omega_{\rm tot}=1$).}
\label{w0wa}
\end{figure}

One of the major promises of the cosmological use of GRBs is related
to the possibility to study the nature of Dark Energy with such a
class of ``standard candles'' extending out to very large redshifts.
With the present sample of 19 GRBs we can explore the equation of
state (EOS) of DE, which can be parametrized in different ways. Given
the already considerably large dispersion of GRB redshifts
(i.e. between 0.168 to 3.2 for the 19 GRBs of our sample) we adopt the
parametrization proposed by Linder \& Huterer (2005) for the EOS of DE, i.e.
$P=w(z)\rho$, where:
\begin{equation}
w(z)= w_{0} + {w_{a}z \over 1+z}
\label{param}
\end{equation}
With  this assumption  the luminosity  distance, as  derived  from the
Friedmann equations, is
\begin{multline}
d_{L}(z;\Omega_{M},w_{0},w_{a}) =  {c(1+z) \over H_{0}} \int_{0}^{z} dz 
[ \Omega_{M} (1+z)^{3}  \\
+ (1-\Omega_{\rm M})(1+z)^{3+3w_{0}+3w_{a}} 
\exp\left( -3w_{a} {z \over 1+z}\right)]^{-1/2}
\label{lumdist}
\end{multline}
which depends on the ($w_{0}$,$w_{a}$) parameters. Note that
Eq. \ref{lumdist} is derived with the prior of a flat Universe.  

\begin{figure}
\centerline{\psfig{figure=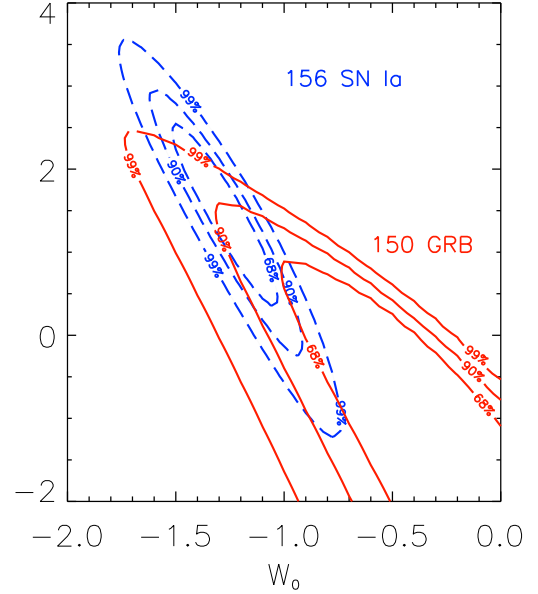,width=9cm}}
\vskip -0.2 true cm
\caption{
Constraints on the cosmological parameters
$w_{0}$, $w_{a}$ obtained with the 150 simulted GRBs (red
contours) compared with the same contours obtained with the 156 SNIa
of the ``Gold'' sample. A flat universe is assumed
($\Omega_{\rm tot}=1$).}
\label{w0wa}
\end{figure}

First we can assume the CMB prior of a flat universe together with the
assumption of a non--evolving equation of state of the Dark Energy
(i.e. $w_{a}=0$). We show in Fig. 5 the contours obtained with the
sample of 150 simulated GRBs and compare with the same constraints
derived with the 156 SN Ia of the ``Gold'' sample. The constraints on
$w_{a}, w_{0}$ are reported in Fig. 6, assuming $\Omega_{\rm M}=0.3$.

\section{The calibration of the spectral energy correlations}

The cosmological use of the $E_{\rm peak}=K\cdot E_{\gamma}^g$
correlation suffers from the so called ``circularity problem''
(Ghirlanda et al. 2004b, Ghisellini et al. 2005): this means that 
both the slope $g$ and the normalization $K$ of the correlation are 
cosmology dependent. In fact, of the two rest frame quantities 
$E_{\rm peak}$ and $E_{\gamma}$ that are used to compute the Ghirlanda 
correlation, the second one ($E_{\gamma}$) depends on the cosmological 
model through the luminosity distance $d_{L}(z;\Omega)$. 
\begin{figure}
\centerline{\psfig{figure=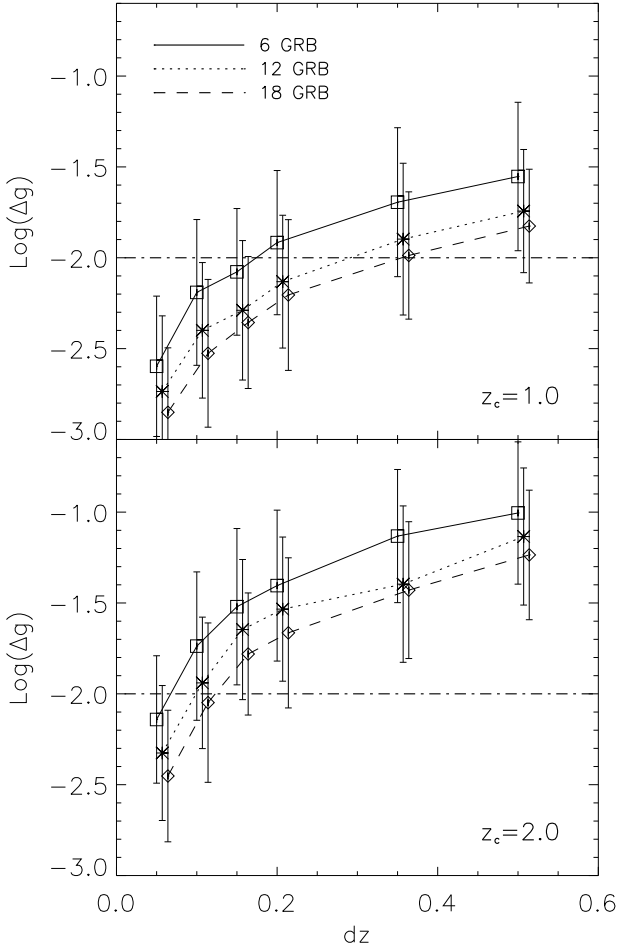,width=9cm}}
\vskip -0.2 true cm
\caption{
Calibration of the $E_{\rm peak}$-$E_{\gamma}^g$ correlation. 
For different samples of GRBs (6, 12, 18 - corresponding to
the solid, dotted and dashed lines respectively) we show the maximum
variation $\Delta g$ of the slope of the correlation for any cosmology
$\Omega\in(0,1.5)$ as a function of the redshift dispersion of the GRBs
$dz$. The dot dashed line represents the limit of variation of 1\% of
the slope of the correlation. Data points have been shifted along the 
abscissa for graphycal purposes.}
\label{calibra} 
\end{figure}
In principle this issue could be solved (a) with a considerably large
number of calibrators, i.e. low redshift GRBs for which the luminosity
distance $d_{L}$ is practically independent from the cosmological
parameters, or (b) with a convincing theoretical interpretation of the 
physical nature of this correlation. In both cases the slope of the 
correlation would be fixed.

Case (a) could be realized with 5--6 GRBs at $z<0.1$. However, if
(long) GRBs are produced by the core--collapse of massive stars, their
rate is mainly regulated by the cosmic SFR and, therefore, the probability of
detecting events at $z<0.1$ is small ($\sim 2\times10^{-5}$). This
number should be convolved with the GRB luminosity function: with the
assumptions (described in Sec. 4) of our simulation we estimate that
$\sim 1.3$\% 
of the 150 GRBs should be at low redshifts
(i.e. $z<0.4$). Instead we should expect to have more chances to detect a 
considerable number (up to $\sim31$\%) of intermediate redshift GRBs
($z\sim 1-2$) where the cosmic SFR peaks. 

For this reason we explore the possibility to calibrate 
the correlation using a sufficient number of GRBs within a small 
redshift bin centered around {\it any} redshift. In fact, if we could 
have a sample of GRBs all at the same redshift the slope of the Ghirlanda 
correlation would be cosmology independent.
Our objective is, therefore, to estimate {\it the minimum number of GRBs 
$N$} within a {\it redshift bin $dz$} centered around a certain {\it
redshift $z_c$} which are required to calibrate the correlation.

In practice the method consists in fitting the correlation for every choice 
of $\Omega$ using a set of $N$ GRBs distributed in the interval $dz$ (centered 
around $z_c$). We consider the correlation to be calibrated (i.e. its slope to 
be cosmology independent) if the change of the slope $g$ is less than 1\%.

The free parameters of this test are the number of GRBs $N$, the ``redshift
slice'' $dz$ and the central value of the redshift distribution $z_c$.
By Monte Carlo technique we use the same sample simulated in Sec. 4
under the WM assumption to minimize the variation of $\Delta
g(\Omega;N,dz,z_c)$ over the $\Omega_{M},\Omega_{\Lambda}\in(0,1.5)$
plane as a function of the free parameters $(N,dz,z_c)$.

We tested different values of $z_{c}$ and different
redshift dispersions $dz\in(0.05,0.5)$. We required a minimum number
of 6 GRBs to fit the correlation in order to have at least 4 degrees
of freedom.  We report our results in Fig. 7. We show the variation of
$\Delta g$ as a function of $dz$ for different samples (6, 12, 18 GRBs -
solid, dotted and dashed curves in Fig. 7). The error bars
show the width of the distribution of the simulation results and not the
uncertainty on the average value. At any redshift the fewer
the number $N$ of GRBs the larger the change of $\Delta g$ (for
the same $dz$) because the correlation is less constrained. The
dependence from $z_c$ is instead different: for larger $z_c$ 
we require a smaller bin $dz$ to keep $\Delta g$ small. 

From the curves reported in Fig. 7 we can conclude that already 12 GRBs
with $z\in(0.9,1.1)$ might be used to calibrate the slope of the
$E_{\rm peak}$-$E_{\gamma}$ correlation. 
At redshift $z_c=2$ instead we
require a smaller redshift bin i.e. $z\in(1.95,1.05)$.  We find that
$N=12$ GRBs with $z\in(0.45,0.75)$ can be used to achieve the same 1\%
precision in the calibration. However, one key ingredient is that the
GRBs used to calibrate the correlation do not have the same peak
energy otherwise they would collapse in one point in the 
$E_{\rm peak}$-$E_{\gamma}$ plane. 
Within the present sample of 19 GRBs 
there are only 4 GRBs within the redshift interval 0.4--0.8
(i.e. 050525, 041006, 020405 and 051022) and 2 of these 
(050525 and 041006) have a very similar $E_{\rm peak}$.

\section{Conclusions}

We have presented the cosmological constraints that can be obtained
with the present sample of 19 GRBs for which all the relevant
quantities, i.e. redshift $z$, peak energy $E_{\rm peak}$ and jet
break time $t_{b}$, has been properly estimated and published in the
literature. 

Following the results of N05, where the $E_{\rm peak}$-$E_{\gamma}$
correlation found under the hypothesis of a uniform jet model with
either a homogeneous (HM) or a wind density profile (WM) were
presented, we derived the constraints on $\Omega_{\rm M}$ and
$\Omega_{\Lambda}$ in these two scenarios. Tighter constraints are
obtained in the WM model. 

We also presented the future of GRBs as cosmological tools. By
simulating a sample of 150 GRBs (which is comparable in number to the
present sample of ``Gold'' SN Ia) we showed that similar (to SN Ia)
tight constraints can be obtained with GRBs either on the present
universe composition and on the nature and evolution of DE
(parametrized with $w_{0}$ and $w_{a}$). We remark that the collection
of such a large sample of GRBs which can be used as standard candles
requires an accurate measurement of their prompt and afterglow
properties. In particular a wide energy spectral coverage is required
to constrain the peak energy and properly compute the bolometric
corrected GRB energetics.

Finally, a large sample of GRBs would help in calibrating the slope of the
$E_{\rm peak}$-$E_{\gamma}$ correlation (in both the HM or WM case). 
To this aim one would require to collect enough low redshift GRBs. However, we
can relax this requirement: we showed that even if not concentrated at
very low $z$, it is sufficient to have a dozen GRBs with a similar redshift
to find the slope of the correlation in a cosmology--independent way
at the level of 1\% accuracy.

\begin{acknowledgements}
We thank A. Celotti and V. Avila--Reese for  useful discussions.  
We thank the Italian MIUR and INAF for founding (Cofin grant 2003020775\_002).
\end{acknowledgements}

\end{document}